\documentclass[journal = cmatex, article, traditional, titlepage, layout=twocolumn]{achemso}

\usepackage{amssymb}
\usepackage{float}
\usepackage{graphicx}
\usepackage{amsmath}
\usepackage{booktabs}
\usepackage{multirow}
\usepackage{array}
\usepackage{makecell}
\usepackage{natbib}
\usepackage{bm}
\usepackage{color}
\usepackage{xcolor}
\usepackage[utf8]{inputenc}
\usepackage[T1]{fontenc}
\usepackage{etoolbox}

\definecolor{darkgreen}{rgb}{0,.6,0}

\author{Andrew\ J.\ E.\ Rowberg}
\affiliation{Materials Department, University of California, Santa Barbara, California 93106-5050, USA}
\author{Michael\ W.\ Swift}
\affiliation{Center for Computational Materials Science, US Naval Research Laboratory, Washington DC 20375, USA}
\altaffiliation{American Society for Engineering Education Postdoctoral Fellow}

\makeatletter
\patchcmd{\acs@contact@details}{E}{*\,E}{}{}
\makeatother

\author{Chris\ G.\ Van de Walle}
\email{vandewalle@mrl.ucsb.edu}
\affiliation{Materials Department, University of California, Santa Barbara, California 93106-5050, USA}

\title{Understanding Carbon Contamination in Proton Conducting Oxides}

\begin{document}

\noindent\textit{Supporting information (SI) for this article is given via a link at the end of the document: Additional data regarding chemical potentials, stability, and defect formation in the cerates and zirconates.}

\begin{abstract}

Carbon contamination is a significant concern for proton-conducting oxides in the cerate and zirconate family, particularly for BaCeO$_3$.
Here, we use first-principles calculations to evaluate carbon stability in SrCeO$_3$, BaCeO$_3$, SrZrO$_3$, and BaZrO$_3$.
The cerates require more carbon-poor environments to prevent carbonate formation, though this requirement can be loosened through the use of more oxygen-poor growth conditions.
Carbonate formation is not the only concern, however.
We find that interstitial carbon has lower formation energies in the cerates relative to the zirconates,
leading to higher carbon concentrations that compete with the desired oxygen vacancy formation.
We also examine the mobility of carbon interstitials, finding that both migration barriers and binding energies to acceptors are lower in the cerates.
As a result, the cerates are likely to degrade when exposed to carbon at operating temperatures.
Our results show definitively why the cerates are less stable than the zirconates with respect to carbon and elucidate the mechanisms contributing to their instability, thereby helping to explain why alloying with zirconium will enhance their operational efficiency.

\end{abstract}




\section{Introduction}
\label{intro}

Considerable research has been devoted to solid-state proton conductors as electrolytes in solid-state hydrogen fuel cells.\cite{SteeleNAT2001}
Two materials in particular have attracted the majority of scientific attention: barium zirconate, BaZrO$_3$ (BZO)\cite{kreuer2003proton,tao2007conductivity} and barium cerate, BaCeO$_3$ (BCO).\cite{iwahara1988bco,katahira2000protonic}
Of these, BCO is often cited for having higher proton conductivities, while BZO is known to be more chemically stable, particularly against CO$_2$.\cite{fabbri2010materials,kochetova2016recent}

Protons are introduced into the zirconates by creating oxygen vacancies during synthesis and then exposing the materials to water, leading to the reaction\cite{kreuer2003proton}:
\begin{equation}
  V_\textrm{O}^{+2}+\textrm{H}_2\textrm{O} \rightarrow 2\textrm{H}_i^+.
\label{Vo+H2O}
\end{equation}
The as-grown material therefore needs to contain a large concentration of oxygen vacancies, which act as electron donors.
This can be accomplished through doping with acceptor impurities, which promote the formation of oxygen vacancies that act as compensating donors.\cite{kreuer2003proton,YamazakiNAT2013,rowberg_zirconates_2019}
However, if another donor species is energetically more favorable than the oxygen vacancy, then acceptor dopants will tend to compensate with that donor instead.
That situation will therefore reduce the efficacy of acceptor doping as an oxygen vacancy generator and, consequentially, limit proton incorporation.
Carbon impurities may compete with oxygen vacancies in this manner.

In addition, carbon can be detrimental because the cerates are prone to decomposing in carbon-rich atmospheres.\cite{iwahara_zirc_1993}
Several studies have shown that fuel cells based on BCO must be operated within a narrow temperature window in order to stave off decomposition when exposed to CO$_2$.\cite{kim2011study,li2014chemical,zakowsky2005elaboration,ryu_stability_1999}
The stability of BCO has been improved demonstrably through alloying with BZO, with approximately equal concentrations of Zr and Ce providing the best balance of stability and conductivity.\cite{sawant2012synthesis,jingde2008chemical,katahira2000protonic,ryu_stability_1999}
Similar results have been demonstrated for Zr-doped SrCeO$_3$ (SCO).\cite{li2008stability}

In previous work, we have shown that the cerates possess poor overall thermodynamic stability, and that SCO has lower stability than BCO,\cite{swift2015small,swift2016impact} which agrees with experimental observations.\cite{scholten1993synthesis,li2008stability}
We have also identified energetically favorable carbon impurity configurations in the zirconates that may compete with the formation of oxygen vacancies.\cite{rowberg_zirconates_2019}
Other studies from our group have investigated similar carbon configurations in other oxides.\cite{lyons2014carbon,tailor_carbon_2015,tailor_impact_2016}
However, we are unaware of any studies aimed at uncovering the atomic-scale reason for these materials' susceptibility to carbon.

Here, we use first-principles techniques based on density functional theory (DFT) with a hybrid functional to study the stability with respect to carbonates, and the incorporation and mobility of carbon species in SCO, BCO, and the equivalent zirconates, SrZrO$_3$ (SZO) and BZO.
We evaluate interstitial C$_i$ and substitutional C$_{\rm \{Ce,Zr\}}$ in all four systems.
We find that, under equivalent conditions for carbon exposure, the formation energy for carbon species in the cerates is considerably lower than in the zirconates, leading to much higher carbon concentrations in the cerates.
We also study carbon impurity migration with the nudged elastic band (NEB) method, along with calculations of binding energies for complexes with the commonly used yttrium acceptor dopant.
Our calculations reveal that carbon mobility is higher in the cerates, explaining why they will suffer from degradation in CO$_2$-rich environments.
Avoiding carbonate-based precursors during synthesis should help limit carbon incorporation.
Doing so is critical, because changing the prevalence of native species---including oxygen---will have no effect unless equilibrium with carbonate formation is broken.
Our results provide new insights into the microscopic reasons for the detrimental effects of carbon in the cerates relative to the zirconates, thereby guiding future research and development on more chemically stable solid-state proton conductors.

\section{Methodology}
\label{method}

\subsection{Computational Details}
\label{comp}

Our calculations are based on DFT within the generalized Kohn-Sham scheme,\cite{kohn_self-consistent_1965} as implemented in the Vienna $Ab$ $initio$ Simulation Package (VASP) \cite{kresse_vasp}.
We use projector augmented wave (PAW) potentials~\cite{blochl_paw1,kresse_paw2} and the hybrid exchange-correlation functional of Heyd, Scuseria, and Ernzerhof (HSE),\cite{heyd_hse} with 25\% mixing of short-range Hartree-Fock exchange.
Consistent with our previous work we use a 400 eV plane-wave cutoff energy for zirconates,\cite{weston_hybrid_2014,weston_acceptor_2017,rowberg_zirconates_2019} and a 500 eV cutoff for cerates.\cite{swift2015small,swift2016impact}
We have confirmed that increasing the cutoff energy for zirconates to 500 eV does not affect our findings.
The Ba $5s^2$ $5p^6$ $6s^2$, Sr $4s^2$ $4p^6$ $5s^2$, Zr $4d^2$ $5s^2$, Ce $5s^2$ $5p^6$ $6s^2$ $5d^1$ $4f^1$, and O $2s^2$ $2p^4$ electrons are treated explicitly as valence.
We model bulk SCO, BCO, SZO, and BZO using orthorhombic unit cells , each containing four formula units, and with a 4$\times$4$\times$3 $k$-point grid to integrate over the Brillouin zone.
BZO has a cubic primitive cell; however, we choose to describe it in an orthorhombic cell to more precisely compare results with the other systems.
To evaluate the energetic barriers associated with defect migration we use the NEB method with climbing images.\cite{henkelman_neb}

\subsection{Defect Calculations}
\label{dcalc}

\begin{figure}
\includegraphics[width=3.5in]{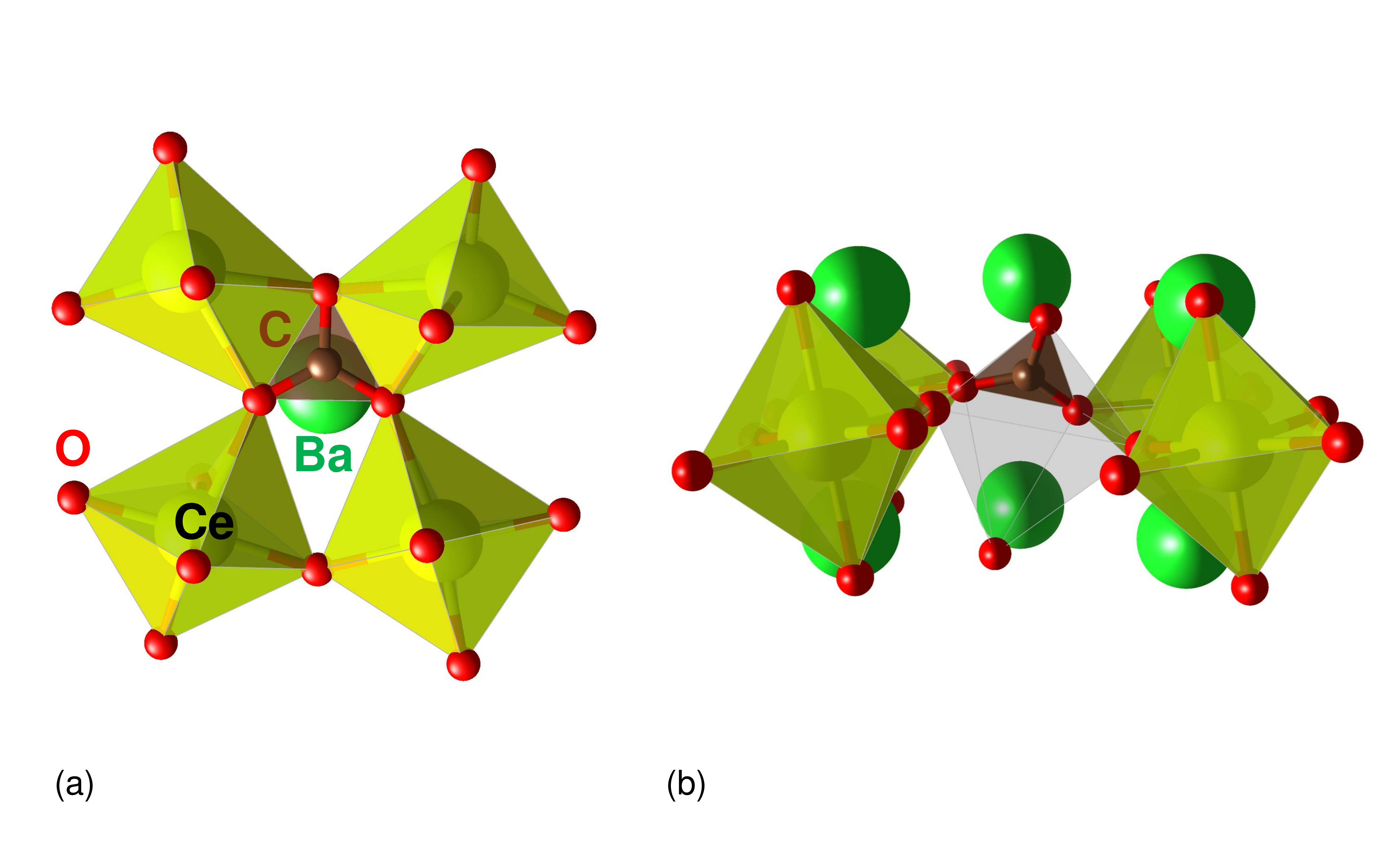}
\caption{Carbon configurations in BaCeO$_3$: (a) interstitial carbon, C$_i$ and (b) substitutional carbon, C$_{\rm Ce}$. Images generated using the VESTA 3 software.\cite{momma2011vesta}}
\label{fig:c-configs}
\centering
\end{figure}

To calculate defect properties, we construct supercells that are $2\times2\times2$ multiples of the orthorhombic unit cell and thus contain 160 atoms.
A $2\times2\times2$ $k$-point grid is used in each case.
We focus specifically on two C configurations in this study: interstitial carbon (C$_i$) and the substitutional species C$_\textrm{\{Ce/Zr\}}$.
Previously, we found these species to be favorable in the zirconates,\cite{rowberg_zirconates_2019} and we assume that they will be similarly favorable in the cerates.
All three configurations are depicted in Fig.~\ref{fig:c-configs} for the case of BCO; the configurations are very similar in the other systems under study.

The formation energy $E^f(D^q)$ of a point defect $D$ in charge state $q$ is calculated as:\cite{FreysoldtRMP2014}
\begin{equation}
\label{eq1}
   E^f(D^q) = E(D^q) - E_{\textrm{bulk}} + \sum n_i\mu_i + qE_F + \Delta_{\textrm{corr}} .
\end{equation}
\noindent $E(D^q)$ is the total energy of a supercell containing defect $D$ in charge state $q$; $E_{\textrm{bulk}}$ is the total energy of a defect-free supercell; $|n_i|$ is the number of atoms of species $i$ added ($n_i<0$) or removed ($n_i>0$) from the system; $\mu_i$ is the chemical potential of species $i$; $E_F$ is the Fermi level referenced to the valence-band maximum (VBM); and $\Delta_{\textrm{corr}}$ is a finite-size correction term.\cite{freysoldt_2009,freysoldt_2011}
The formation energy determines the defect concentration $c$ via a Boltzmann relation:
\begin{equation}
  c = N_{\rm sites} \exp \left( -\frac{E^f}{k_BT}\right) \, ,
\label{eq:conc}
\end{equation}
meaning that lower formation energies result in higher defect concentrations.
$N_{\rm sites}$ is the number of sites (per unit volume) on which the defect can incorporate.
In standard defect formation energy diagrams, $E_F$ is the free variable and varies freely from the VBM to the conduction-band minimum (CBM).
Because of the exponential relationship between formation energy and defect concentration the actual position of the Fermi level will, to a good approximation, be determined by the intersection of the lowest-energy negatively and positively charged defects, which together ensure charge neutrality.

The chemical potentials $\mu_i$ are variables that reflect the abundance of specific elements during synthesis.
We express them in terms of deviations $\Delta \mu_i$ from the total energies of the elemental reference structures, i.e., the ground-state structures of the Sr, Ba, Zr, or Ce metals, or an O atom in O$_2$.
Assuming conditions close to equilibrium, the $\Delta \mu_i$ are related by:
\begin{multline}
\label{eq:stability}
\Delta \mu_\mathrm{\{Sr,Ba\}} + \Delta \mu_\mathrm{\{Ce,Zr\}} + 3\Delta \mu_\mathrm{O}  = \\ \Delta H^f(\textrm{\{Sr,Ba\}\{Ce,Zr\}O}_3) ,
\end{multline}
\noindent where $\Delta H^f(\textrm{\{Sr,Ba\}\{Ce,Zr\}O}_3)$ is the enthalpy of formation for \{Sr,Ba\}\{Ce,Zr\}O$_3$.
Our calculated enthalpies of formation are listed in Table~\ref{tab:enth}.

\begin{table}
\setlength{\tabcolsep}{5pt}
\setlength{\extrarowheight}{4pt}
\centering
\begin{tabular}{ccc} \hline \hline
Compound & $\Delta H^f$ (eV) (calc) & $\Delta H^f$ (eV) (exp) \\ \hline
SrCeO$_3$ & --16.87 & --17.49\cite{cordfunke1998thermochemical} \\
BaCeO$_3$ & --16.77 & --17.52\cite{cordfunke1998thermochemical} \\
SrZrO$_3$ & --17.38 & --18.28\cite{HUNTELAAR19941095} \\
BaZrO$_3$ & --17.29 & --18.28\cite{HUNTELAAR19941095} \\
SrCO$_3$  & --11.98 & --12.65\cite{lide2012crc} \\
BaCO$_3$  & --11.91 & --12.58\cite{lide2012crc} \\
SrO & --5.61 & --6.14\cite{lide2012crc} \\
BaO & --5.09 & --5.68\cite{lide2012crc} \\
ZrO$_2$ & --10.99 & --11.41\cite{lide2012crc} \\
CeO$_2$ & --11.29 & --11.29\cite{lide2012crc} \\
Y$_2$O$_3$ & --19.05 & --19.75\cite{lide2012crc} \\\hline \hline
\end{tabular}
\caption{Calculated and reported enthalpies of formation (in eV per formula unit) for compounds pertinent to this study.}
\label{tab:enth}
\end{table}

For the purposes of presenting our results, we will use $\Delta \mu_\textrm{O}$=$-2.42$ eV, as we have done in previous studies, which is representative of sintering SZO in air at 1650 $^\circ$C.\cite{yajima1992protonic,weston_acceptor_2017,rowberg_zirconates_2019}
These conditions favor $V_\textrm{O}$ formation, as is desired in device applications.
Once $\Delta \mu_\textrm{O}$ is chosen, eq~\ref{eq:stability} fixes the sum of $\Delta \mu_\textrm{\{Sr,Ba\}}$ and $\Delta \mu_\textrm{\{Ce,Zr\}}$.
We focus on \{Sr,Ba\}-poor (\{Ce,Zr\}-rich) values, where $\Delta \mu_\textrm{\{Ce,Zr\}}$ is maximized through equilibrium with \{Ce,Zr\}O$_2$.
These conditions lead to higher carbon concentrations and therefore reflect a worst-case scenario at the chosen oxygen chemical potential.
Choosing a specific set of chemical potential conditions is important for purposes of presenting our results; however, other conditions can easily be examined by referring to eq~\ref{eq1}.
Importantly, our results comparing relative formation energies are not affected by the particular choice of chemical potentials.

We focus on yttrium (Y$_{\rm \{Ce,Zr\}}^-$) as the acceptor dopant, since it is the most commonly used dopant in both the cerates and the zirconates.\cite{kreuer2003proton}
Chemical potentials for Y are chosen at the stability limit for our compounds relative to Y$_2$O$_3$:
\begin{equation}
\label{eq:limiting}
2\Delta \mu_\mathrm{Y} + 3\Delta \mu_\mathrm{O} \leq \Delta H^f(\textrm{Y}_2{\rm O}_3) .
\end{equation}

For carbon, we consider values for $\Delta \mu_{\rm C}$ corresponding to the solubility limit by referring to the limiting condition of carbonate formation:
\begin{equation}
\label{eq:carbonate}
\Delta \mu_\mathrm{\{Sr,Ba\}} + \Delta \mu_{\rm C} + 3\Delta \mu_\mathrm{O} \leq \Delta H^f(\textrm{\{Sr,Ba\}C}{\rm O}_3) .
\end{equation}
Considering that SrCO$_3$ and BaCO$_3$ are often used as precursors in synthesis of the cerates and zirconates,\cite{ryu_stability_1999,iwahara1988bco,iwahara_zirc_1993} this limiting condition is appropriate.
For higher values of $\Delta\mu_{\rm C}$, the carbonate phases will form preferentially to the zirconates and cerates;
thus, for any choice of $\Delta \mu_{\rm \{Sr,Ba\}}$ and $\Delta \mu_{\rm O}$, this condition determines the maximum allowed carbon chemical potential.
For more oxygen-rich conditions, the limiting value for $\Delta\mu_{\rm C}$ will decrease; as a result, more oxygen-rich conditions require lower carbon chemical potentials (less carbon in the environment) to avoid destabilizing the zirconates and cerates against their respective carbonate phases.

\section{Results and Discussion}
\label{res}

\subsection{Properties of Point Defects and Impurities}
\label{defect}

In order to plot defect formation energies, we must choose specific chemical potential conditions.
As discussed, we focus on the \{Sr,Ba\}-poor limit, where $\Delta \mu_{\rm C}$ is maximized (``worst case'' conditions).
In Table~\ref{tab:Cchempot}, we list the maximum values of $\Delta \mu_{\rm C}$ for $\Delta \mu_{\rm O}=-2.42$ eV, which reflects typical synthesis conditions, and for $\Delta \mu_{\rm O}=-1$ eV, which represents more O-rich conditions (we include the full chemical potentials in Tables S1 and S2 in the SI).
Values of $\Delta \mu_{\rm C}$ will vary linearly within this range.
The values in Table~\ref{tab:Cchempot} show that greater values of $\Delta \mu_{\rm C}$ are permitted in the zirconates as compared to the cerates, which means that the zirconates are stable with respect to carbonates under a broader range of environments.
Moving to more O-poor synthesis (i.e., lowering $\Delta \mu_{\rm O}$) will increase this limit and, correspondingly, the permissible range of carbon conditions.

\begin{table}
\setlength{\tabcolsep}{5pt}
\setlength{\extrarowheight}{4pt}
\centering
\begin{tabular}{ccc} \hline \hline
\thead{Compound} & \thead{$\Delta \mu_{\rm C}$ (eV) \\ ($\Delta \mu_{\rm O}=-2.42$ eV)} & \thead{$\Delta \mu_{\rm C}$ (eV) \\ ($\Delta \mu_{\rm O}=-1$ eV)} \\ \hline
SrCeO$_3$ & --1.43 & --4.27 \\
BaCeO$_3$ & --1.47 & --4.30 \\
SrZrO$_3$ & --0.75 & --3.59 \\
BaZrO$_3$ & --0.77 & --3.61 \\ \hline \hline
\end{tabular}
\caption{Maximum carbon chemical potentials (at the \{Sr,Ba\}-poor limit) in the cerates and zirconates under selected oxygen chemical potential conditions.}
\label{tab:Cchempot}
\end{table}

\begin{figure*}
\includegraphics[width=5.5in]{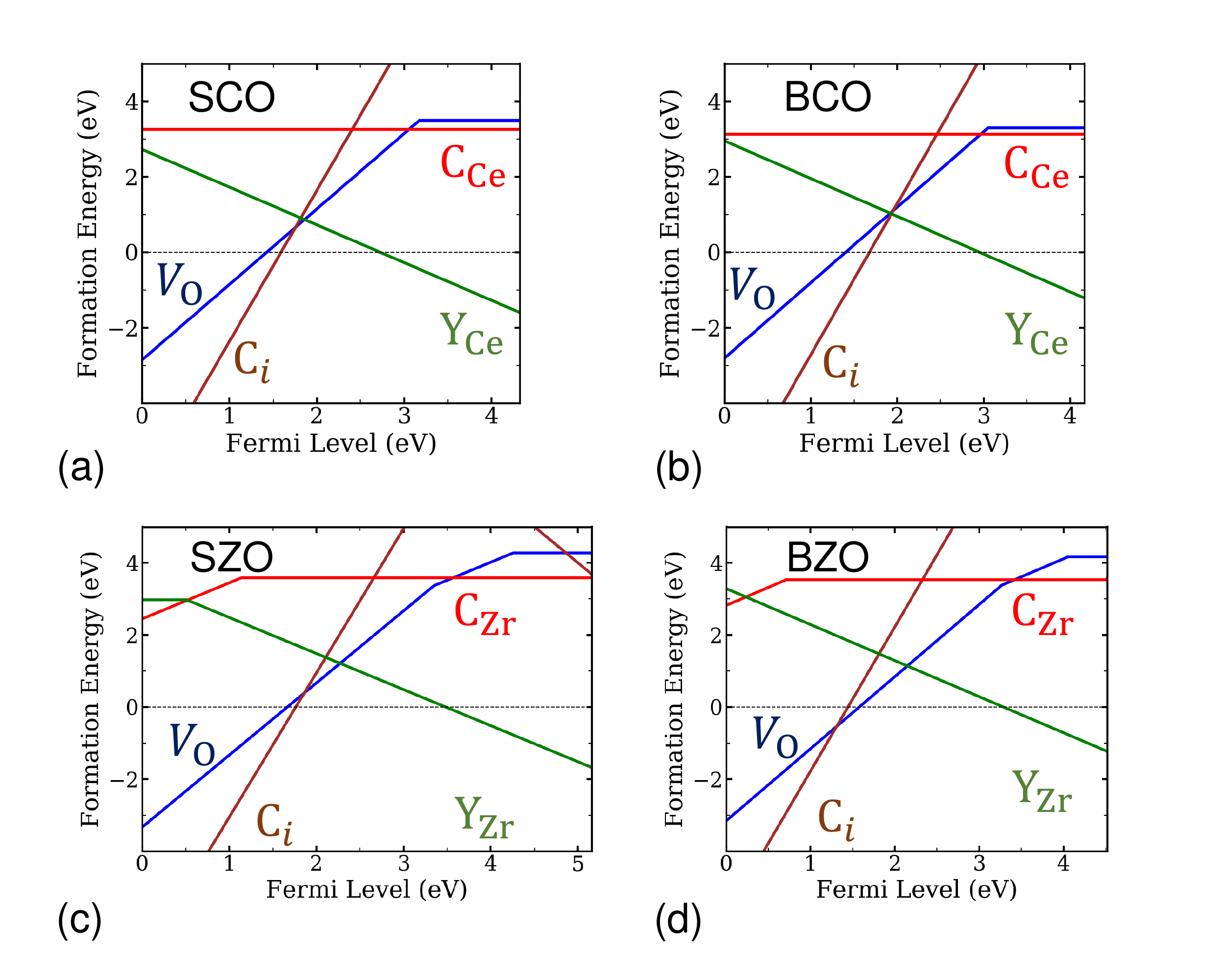}
\caption{Formation energies for oxygen vacancies, yttrium dopants, and carbon impurities in (a) SrCeO$_3$, (b) BaCeO$_3$, (c) SrZrO$_3$, and (d) BaZrO$_3$ under C-rich, \{Sr,Ba\}-poor conditions, and with $\Delta \mu_{\rm O}=-2.42$ eV.}
\label{fig:def}
\centering
\end{figure*}

Using the chemical potential conditions identified in Table~\ref{tab:Cchempot} for $\Delta \mu_{\rm O}=-2.42$ eV, we plot defect formation energies for C$_i$, C$_{\rm \{Ce,Zr\}}$, $V_{\rm O}$, and Y$_{\rm \{Ce,Zr\}}$ in the cerates and zirconates in Fig.~\ref{fig:def}.
Note that results for other chemical potentials can readily be obtained by referring to eq~\ref{eq1}.
As we previously discovered for the zirconates,\cite{rowberg_zirconates_2019} C$_i^{+4}$ is a very favorable carbon configuration, while C$_\textrm{Ce}^0$ is higher in energy.
In each case, carbon atoms bond with three neighboring oxygen atoms, with C--O bond lengths ranging from 1.28 \AA \: to 1.30 \AA, which matches the geometry of the CO$_3^{2-}$ ion (see Fig.~\ref{fig:c-configs}).

We turn our attention to the Fermi level corresponding to charge neutrality in each system.
For SCO, SZO and BZO, charge neutrality will be determined by compensation between Y$_{\rm \{Ce,Zr\}}^-$ and $V_{\rm O}^{+2}$.
In BCO, C$_i^{+4}$ will actually be the compensating donor species, and in SCO, the formation energy of C$_i^{+4}$ is very close to that of $V_{\rm O}^{+2}$ at the position of charge neutrality, meaning that a large concentration of C$_i^{+4}$ will be present.
In SZO and BZO, C$_i^{+4}$ has a much higher formation energy at the Fermi level corresponding to charge neutrality.
The C$_i^{+4}$ species in BZO has a particularly high formation energy, helping to explain why low-energy acceptor dopants will not promote carbon contamination.\cite{kang2017first,polfus_co2-bzo_2018}
Thus, we expect the cerates to contain higher concentrations of C$_i$ species, even when carbonate formation is not thermodynamically favored.
Dopants with lower formation energies than Y$_{\rm Ce}^-$ would actually compensate with C$_i^{+4}$ in SCO.
Choosing a dopant with a {\it higher} formation energy will decrease the carbon concentration, albeit at the expense of also decreasing the concentration of oxygen vacancies.

Choosing more carbon-poor conditions---as could be accomplished by avoiding the use of carbonate precursors during synthesis---increases the formation energy (and hence decreases the concentration) of both carbon species relative to the other defects.
Changing the chemical potentials of host species (\{Sr,Ba\}, \{Ce,Zr\}, or O), however, will not qualitatively affect our observations.
Indeed, we have checked that formation energies shift only modestly (by at most 0.2--0.3 eV) when moving from \{Sr,Ba\}-rich to \{Sr,Ba\}-poor conditions.
This finding is directly related to the narrow windows of chemical stability in each compound, which are particularly narrow for the cerates (see Fig. S1 in the SI).
Furthermore, while changing $\Delta \mu_{\rm O}$ moves each individual formation energy line, it does not change the formation energies of species at the charge neutrality point; it simply has the net effect of shifting the position of the Fermi level (see Fig. S2 in the SI).

This point is important to emphasize: carbon incorporation will not be affected by changing native chemical potentials unless equilibrium with the carbonate species is broken.
Using a synthesis route that does not involve carbonate phases is thus essential to limit carbon incorporation.
One possible method involves the use of nitrate precursors [Sr(NO$_3$)$_2$, Ba(NO$_3$)$_2$, Ce(NO$_3$)$_3$, and ZrO(NO$_3$)$_2$], which have previously been used for synthesis of both cerates\cite{chen1997preparation,flint1995comparison} and zirconates\cite{potdar2000preparation,suresh2010synthesis}.
We previously found nitrogen impurities to have high formation energies in the zirconates,\cite{rowberg_zirconates_2019} which, following from the results we have just presented for carbon impurities, suggests that they will also have high formation energies in the cerates.
As a result, nitrate precursors would be less deleterious for chemical stability.

\subsection{Carbon Migration}
\label{mig}

\begin{figure}
\includegraphics[width=3.5in]{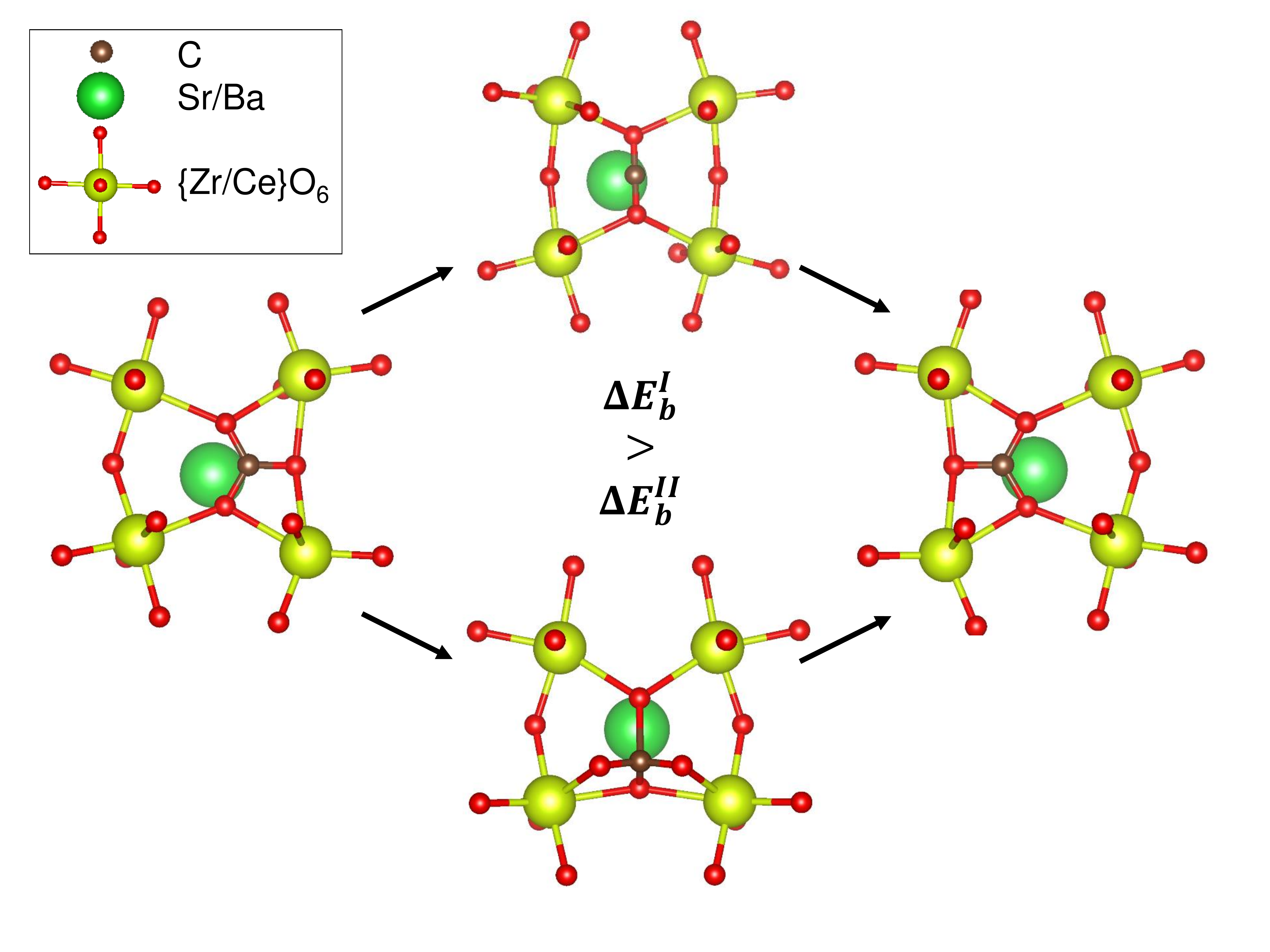}
\caption{Migration pathway for the carbon interstitial (C$_i^{+4}$) in the cerates and zirconates. The top pathway has a higher energetic barrier than the lower pathway.}
\label{fig:c-path}
\centering
\end{figure}

To further understand carbon contamination, we study carbon mobility in the cerates and zirconates.
To do so, we calculate migration barriers for carbon motion using the NEB method.
The carbon interstitial will be the most mobile species.
C$_i^{+4}$ typically prefers to be nestled between two \{Ce,Zr\}--O$_6$ octahedra, which share one oxygen atom to which a C--O bond is formed.
The other two C--O bonds connect C$_i^{+4}$ with one oxygen atom in each of the octahedra (left panel of Fig.~\ref{fig:c-path}).
We investigate two primary pathways for C$_i^{+4}$ migration, which
differ in the way in which the C$_i^{+4}$ species can get around these nearby oxygen atoms: passing directly between them in a straight line, or swinging around them by moving out-of-plane.
These two possible pathways are shown schematically in Fig.~\ref{fig:c-path}.
We find that the ``swinging'' mechanism (the lower pathway in Fig.~\ref{fig:c-path}) is much more favorable, even though several new C--O bonds are formed and broken during the process.
In the higher-energy ``passing-between'' pathway (the upper pathway of Fig.~\ref{fig:c-path}), the saddle-point configuration forces carbon to adopt a two-fold coordination with oxygen, which is unfavorable in light of carbon's energetic preference to be coordinated with at least three oxygen atoms.

In Table~\ref{tab:migration}, we list migration barriers for the lowest-energy pathways we have calculated for C$_i$ in each material.
For materials sharing an A--site cation, the zirconates have larger barriers for migration.
It has been suggested that the cubic symmetry of BZO is related to its stability with respect to carbon;\cite{sawant2012synthesis} however, it is clear that migration barriers alone do not bear out that supposition.
The migration barrier for BZO is slightly larger than that of the other materials, but its barrier relative to SZO is in line with the same trend in the cerates, making it difficult to point to symmetry as a determining factor.
Overall, the barriers are such that we expect C$_i^{+4}$ to be mobile during synthesis, and possibly also during device operation, which is typically at temperatures on the order of 1000 K.
\begin{table}
\setlength{\tabcolsep}{5pt}
\setlength{\extrarowheight}{4.5pt}
\caption{Calculated migration barriers $E_b$ and binding energies (with the yttrium acceptor dopant) $E_{\rm bind}$ for carbon interstitials in proton-conducting oxides.}
\label{tab:migration}
\centering
\begin{tabular}{lcc} \hline \hline
  \thead{Material} & \thead{C$_i^{+4}$ (eV)} & \thead{$E_{\rm bind}$[C$_i^{+4}$- Y$_{\rm \{Ce,Zr\}}^-$] (eV)} \\ \hline
  SrCeO$_3$ & 1.03  & 0.52 \\
  BaCeO$_3$ & 2.13  & 0.60 \\
  SrZrO$_3$ & 1.87  & 0.95 \\
  BaZrO$_3$ & 2.39  & 0.98 \\ \hline \hline
\end{tabular}
\end{table}

An additional important consideration for mobility is the binding energy of complexes between acceptor dopants and carbon impurities.
As acceptors like Y$_{\rm Zr}^-$ have an opposite charge to C$_i^{+4}$, there will be a Coulombic binding energy hindering the movement of carbon in highly doped samples.
Dopants may in fact trap carbon impurities as they move within the material, potentially decelerating decomposition reactions.

We calculate the binding energy between an acceptor $A^-$ and a donor $D^+$ as:
\begin{equation}
  E_{\rm bind}(AD)=E^f(A^-)+E^f(D^+)-E^f(AD) ,
\label{eq:bind}
\end{equation}
where $E^f(A^-)$ and $E^f(D^+)$ are formation energies of isolated defects and $E^f(AD)$ is the formation energy of a complex containing both defects in close proximity in the same simulation cell.
Practically speaking, $E_{\rm bind}$ is a dissociation energy needed to break apart defect complexes.
Given that C$_i^{+4}$ is the most significant mobile defect species, we calculate its binding energy with yttrium acceptors in the cerates and zirconates.
These energies will need to be overcome, in addition to the migration barriers previously calculated, for carbon to be mobile.

Our calculated binding energies are listed in the second column of Table~\ref{tab:migration}.
Clearly, binding energies are significantly smaller in the cerates than in the zirconates, which, coupled with our results for migration barriers, further demonstrates that carbon will be more mobile in the cerates.
Combining these binding energies with the calculated migration barriers, we can conclude that carbon will be largely immobile in the zirconates at the operating temperature, but can relatively easily penetrate the cerates.
Alloying the cerates with the zirconates will increase binding energies and migration barriers, which helps to explain the better stability observed in alloyed materials.\cite{kochetova2016recent,sawant2012synthesis,jingde2008chemical,li2008stability,ryu_stability_1999}

\section{Conclusions}
\label{conc}

We have examined the formation and mobility of carbon impurity species in the cerates and zirconates.
We show that the cerates have lower chemical stability with respect to carbonates at specific oxygen chemical potentials, and that synthesis at more oxygen-poor conditions will permit a wider range of carbon chemical potentials while avoiding carbonate formation.
But carbonate formation is not the only concern:
we find that interstitial C$_i^{+4}$ can incorporate in both the cerates and the zirconates, and it competes with $V_{\rm O}^{+2}$, thus hindering prospects for proton conduction.
Carbon contamination is particularly severe for the cerates, because of both the lower formation energy of C$_i$ as well as the higher mobility of interstitial carbon, which allows carbon to penetrate into the cerates at typical operating temperatures.
Choosing dopants other than yttrium can suppress carbon incorporation, but at the expense of lowering $V_{\rm O}$ concentrations.

We also find that simply changing host-atom chemical potentials during synthesis cannot suppress carbon incorporation, and thus avoiding exposure to carbon is essential for the cerates; use of precursors other than carbonates could be a promising route.
Exposure to carbon during operation will also be detrimental for cerates.
Overall, the zirconates are less sensitive to all of these deleterious effects, and therefore, alloying cerates with zirconates will improve stability.
For pure cerates, completely avoiding carbon contamination will be crucial for unlocking their performance in proton-conducting applications.

\section*{Conflicts of interest}
There are no conflicts to declare.

\section*{Acknowledgements}
A.J.E.R. was supported by the National Science Foundation (NSF) Graduate Research Fellowship Program under Grant No. 1650114. Any opinions, findings, and conclusions or recommendations expressed in this material are those of the author(s) and do not necessarily reflect the views of the NSF. M.W.S. acknowledges support from the Naval Research Laboratory Postdoctoral Fellowship through the American Society for Engineering Education.  C.G.V.d.W. was supported by the Office of Science of the U.S. Department of Energy (DOE) (Grant No. DE-FG02-07ER46434). We acknowledge the use of the Center for Scientific Computing supported by the California NanoSystems Institute and the Materials Research Science and Engineering Center (MRSEC) at UC Santa Barbara through NSF DMR 1720256 and NSF CNS 0960316. We also acknowledge computational resources provided through the National Energy Research Scientific Computing Center, a DOE Office of Science User Facility supported by the Office of Science of the U.S. DOE under Contract No. DE-AC02-05CH11231.

\clearpage
\bibliographystyle{apsrev}
\bibliography{PCOBib}

\end{document}


\section*{S1: Chemical Potential Tables}

As discussed in the main text, we consider two selected oxygen chemical potential conditions: $\Delta\mu_{\rm O}=-2.42$ eV, which corresponds to experimental sintering conditions,\cite{yajima1992protonic} and $\Delta\mu_{\rm O}=-1$ eV, which represents more O-rich conditions.
Table 2 in the main text lists the maximum allowed carbon chemical potentials under both of these conditions, assuming Sr/Ba-poor synthesis.
In Table~\ref{tab:chempot_O242}, we list the full chemical potentials ($\Delta\mu_{\rm \{Sr,Ba\}}$, $\Delta\mu_{\rm \{Ce,Zr\}}$ and $\Delta\mu_{\rm C}$) for the cerates and zirconates for $\Delta\mu_{\rm O}=-2.42$ eV.
In Table~\ref{tab:chempot_O1}, we do the same for $\Delta\mu_{\rm O}=-1$ eV.

\begin{table}[h]
\setlength{\tabcolsep}{5pt}
\setlength{\extrarowheight}{4pt}
\centering
\begin{tabular}{cccc} \hline \hline
\thead{Compound} & \thead{$\Delta \mu_{\rm \{Sr,Ba\}}$ (eV)} & \thead{$\Delta \mu_{\rm \{Ce,Zr\}}$ (eV)} & \thead{$\Delta \mu_{\rm C}$ (eV)} \\ \hline
SrCeO$_3$ & --3.29 & --6.45 & --1.43 \\
BaCeO$_3$ & --3.18 & --6.45 & --1.47 \\
SrZrO$_3$ & --3.97 & --6.15 & --0.75 \\
BaZrO$_3$ & --3.88 & --6.15 & --0.77 \\ \hline \hline
\end{tabular}
\caption{Chemical potentials in the cerates and zirconates at C-rich, Sr/Ba-poor conditions, with $\Delta\mu_{\rm O}=-2.42$ eV.}
\label{tab:chempot_O242}
\end{table}

\begin{table}[h]
\setlength{\tabcolsep}{5pt}
\setlength{\extrarowheight}{4pt}
\centering
\begin{tabular}{cccc} \hline \hline
\thead{Compound} & \thead{$\Delta \mu_{\rm \{Sr,Ba\}}$ (eV)} & \thead{$\Delta \mu_{\rm \{Ce,Zr\}}$ (eV)} & \thead{$\Delta \mu_{\rm C}$ (eV)} \\ \hline
SrCeO$_3$ & --4.71 & --9.29 & --4.27 \\
BaCeO$_3$ & --4.60 & --9.29 & --4.31 \\
SrZrO$_3$ & --5.39 & --8.99 & --3.59 \\
BaZrO$_3$ & --5.30 & --8.99 & --3.61 \\ \hline \hline
\end{tabular}
\caption{Chemical potentials in the cerates and zirconates at C-rich, Sr/Ba-poor conditions, with $\Delta\mu_{\rm O}=-1$ eV.}
\label{tab:chempot_O1}
\end{table}

\section*{S2: Chemical Stability Diagrams}

\begin{figure}[h]
\includegraphics[width=13.5cm]{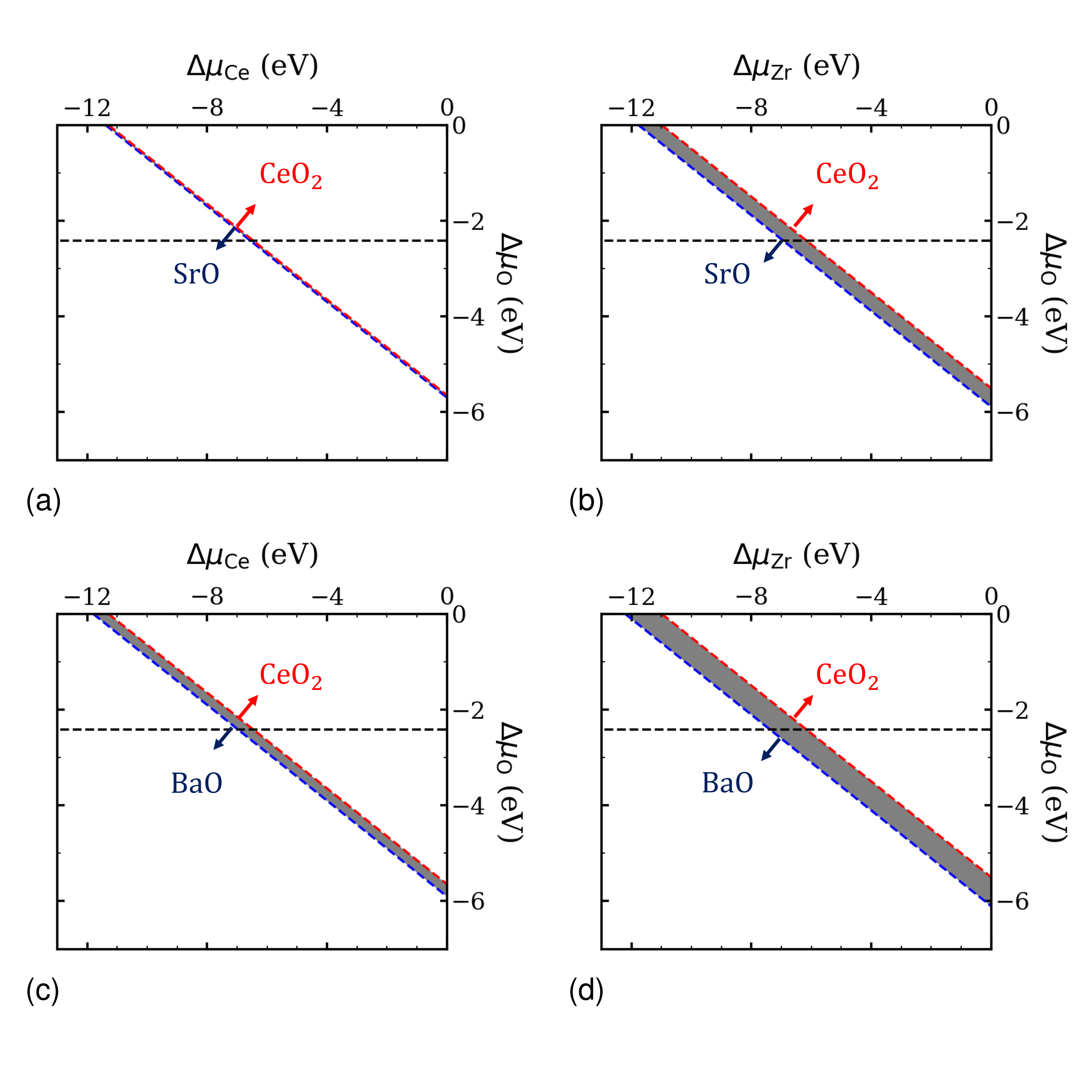}
\caption{Chemical potential phase diagrams for (a) SrCeO$_3$, (b) SrZrO$_3$, (c) BaCeO$_3$, (d) BaZrO$_3$. Competing phases are indicated.}
\label{fig:stability}
\end{figure}

We plot the chemical stability regions (shaded in gray) for the cerates and zirconates in Fig.~\ref{fig:stability}.
The stability regions for each cerate and zirconate are narrow in $\Delta\mu_{\rm \{Ce,Zr\}}$-vs.-$\Delta\mu_{\rm O}$ space.
For each compound, the widths are as follows: 0.05 eV for SCO [Fig.~\ref{fig:stability}(a)], 0.39 eV for SZO [Fig.~\ref{fig:stability}(b)], 0.26 eV for BCO [Fig.~\ref{fig:stability}(c)], and 0.61 eV for BZO [Fig.~\ref{fig:stability}(d)].
The narrow widths explain the low variability in formation energies when comparing the Sr/Ba-rich and Sr/Ba-poor limits.
Note that Refs.~\cite{swift2015small,swift2016impact} considered additional limiting phases for SCO and BCO; however, for the oxygen chemical potentials of interest, the regions where those phases form are not relevant, so we neglect them here.

\section*{S3: Effect of Changing $\Delta\mu_{\rm O}$ on Formation Energy Diagrams}

\begin{figure}[h]
\includegraphics[width=12cm]{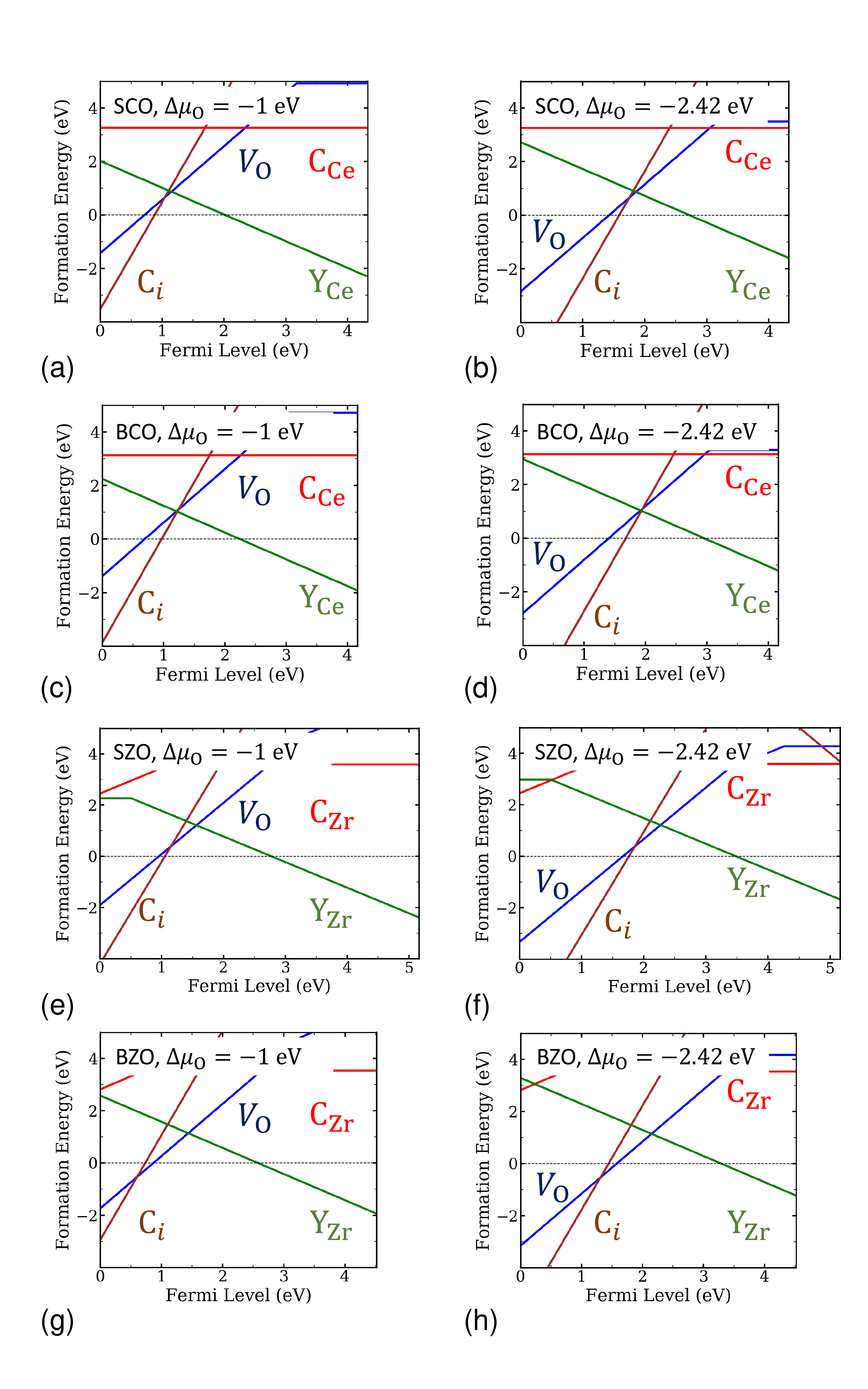}
\caption{Defect formation energies for SCO under Sr-poor conditions for (a) $\Delta\mu_{\rm O}=-1$ eV and (b) $\Delta\mu_{\rm O}=-2.42$ eV; defect formation energies for BCO under Ba-poor conditions for (c) $\Delta\mu_{\rm O}=-1$ eV and (c) $\Delta\mu_{\rm O}=-2.42$ eV; defect formation energies for SZO under Sr-poor conditions for (e) $\Delta\mu_{\rm O}=-1$ eV and (f) $\Delta\mu_{\rm O}=-2.42$ eV; and defect formation energies for BZO under Ba-poor conditions for (g) $\Delta\mu_{\rm O}=-1$ eV and (h) $\Delta\mu_{\rm O}=-2.42$ eV.}
\label{fig:muO-vary}
\end{figure}

As discussed in the main text, changing the oxygen chemical potential does not change the formation energy at the charge neutrality point; instead, it merely shifts the Fermi level at which charge neutrality is achieved to lower energies.
We show this trend graphically for each compound in Fig.~\ref{fig:muO-vary}, with results for $\Delta\mu_{\rm O}=-1$ eV in the left column and those for $\Delta\mu_{\rm O}=-2.42$ eV (also shown in Fig. 2 in the main text) on the right.
Sr/Ba-poor conditions are used throughout.

\clearpage
\bibliographystyle{apsrev}
\bibliography{PCOBib}